\documentclass[pdflatex,iicol,sn-mathphys-num]{sn-jnl}
\usepackage{graphicx,amsmath,amssymb,xcolor,array}
\newenvironment{acknowledgments}{\section*{ACKNOWLEDGMENTS}}{\par}
\hypersetup{pdfauthor={Kilar Zhang, Alessandro Parisi, C. Vásquez Flores, C. Henrique Lenzi, Chian-Shu Chen},
  pdftitle={Compact stars with the same mass, tidal deformability and f-mode frequency}}
\newcommand{\ms}{M_{\odot}}
\newcommand{\dd}{\mathrm{d}}
\newcommand{\Ltil}{\widetilde\Lambda}

\renewcommand{\addressfont}{\normalfont\fontsize{9bp}{10.5bp}\selectfont\titraggedcenter}
\begin{document}
\title{Compact stars with the same mass, tidal deformability and $f$-mode frequency}

\author[1,2,3]{\fnm{Kilar} \sur{Zhang}}\email{kilar@shu.edu.cn}
\author[4,5]{\fnm{Alessandro} \sur{Parisi}}\email{alessandro.parisi@unipg.it}
\author[6,7]{\fnm{C.} \sur{V\'asquez Flores}}\email{cesar.vasquez@uemasul.edu.br}
\author[8]{\fnm{C. Henrique} \sur{Lenzi}}\email{chlenzi1980@gmail.com}
\author[9,10]{\fnm{Chian-Shu} \sur{Chen}}\email{chianshu@gmail.com}

\affil[1]{Department of Physics and Institute for Quantum Science and Technology, Shanghai University, Shanghai 200444, Mainland China}
\affil[2]{Shanghai Key Lab for Astrophysics, Shanghai 200234, Mainland China}
\affil[3]{Shanghai Key Laboratory of High Temperature Superconductors, Shanghai 200444, Mainland China}
\affil[4]{Department of Physics, Perugia University, Italy}
\affil[5]{INFN, Sezione di Perugia, Via Alessandro Pascoli 23c, 06123 Perugia, Italy}
\affil[6]{Centro de Ci\^encias Exatas, Naturais e Tecnol\'ogicas, UEMASUL, Rua Godofredo Viana 1300, CEP 65901-480, Imperatriz, Maranh\~ao, Brazil}
\affil[7]{Departamento de F\'isica, CCET, Universidade Federal do Maranh\~ao, Campus Universitario do Bacanga, CEP 65080-805, S\~ao Lu\'is, MA, Brasil}
\affil[8]{Departamento de F\'isica, Instituto Tecnol\'ogico de Aeron\'autica, DCTA, 12228-900 S\~ao Jos\'e dos Campos, SP, Brazil}
\affil[9]{Department of Physics, Tamkang University, New Taipei 251, Taiwan}
\affil[10]{Physics Division, National Center for Theoretical Sciences, Taipei 10617, Taiwan}

\abstract{Simultaneous measurements of the quadrupolar tidal deformability $\Lambda_2$ and the fundamental-mode frequency $f_2$ of a compact star have been proposed as a way to identify self-bound strange stars. We test this proposal with explicit counterexamples in full general relativity. Starting from a calibrated analytic neutron-star barotrope, we tune the bag constant of strange quark matter, realized inside the bulk stability window and supporting $2\,\ms$, until the quark star and the neutron star share their mass, $\Lambda_2$ and the real part of $f_2$. For massless quarks the coincidence occurs at $M\simeq1.50\,\ms$, $\Lambda_2\simeq364$ and $f_2\simeq1.733$~kHz. It persists when a 95~MeV strange-quark mass, a realistic SLy crust or a frozen-composition adiabatic index is included, although its location in $\Lambda_2$ moves by up to 18\%. More generally, at fixed mass and $\Lambda_2$ the frequencies of all our models that support $2\,\ms$ agree within about 0.5~Hz, below the 1.3~Hz linewidth of the mode and far below the projected measurement precision, so $f_2$ adds essentially no information to $(M,\Lambda_2)$. The coincident stars differ instead in radius, by 0.2~km for the analytic surface and 0.6--0.8~km once a crust is attached, and in their response at other masses: at $1.85\,\ms$ the same two equations of state differ by 15\% in $\Lambda_2$ and by 56~Hz in $f_2$. In binaries they define a curve of equal leading tidal parameter, on which the remaining dephasing could be resolved only at signal-to-noise ratios above $4\times10^4$, even with all masses known.}
\onecolumn
\maketitle
\twocolumn

\section{Introduction}\label{sec:intro}
%=====================================================================
The fundamental ($f$) mode of a compact star couples more strongly than any other oscillation to gravitational waves and to the tidal field of a binary companion. Its frequency correlates with the mean density~\cite{ak}, and in general relativity the mass-scaled frequency $M\omega_\ell$ is an almost equation-of-state (EoS) independent function of the tidal deformability $\Lambda_\ell$ of the same multipole~\cite{chan}. This $f$--Love relation and its relatives involving compactness or the moment of inertia~\cite{lau2010,chirenti2015,sotani2021,yagiyunes2017} have been tested on increasingly broad EoS samples~\cite{zhao,chu}. They hold for self-bound quark stars as well as for neutron stars~\cite{chan,zhao}, and a recent full-GR survey finds that the residuals of the $f$--$\Lambda$ relations are nearly uncorrelated with the EoS~\cite{chu}. Universality is what makes the $f$ mode useful for inferring radii and tidal parameters. It also suggests that a measured $f_2$ carries little information that the mass and $\Lambda_2$ do not already contain.

Whether a compact star is self-bound, i.e.\ has a finite surface density, is a long-standing question~\cite{witten,afo,hzs}. Radius, tidal, inertia and magnetic-Love signatures have been studied~\cite{postnikov,iloveself,fukushima}. Pradhan, Shirke and Chatterjee proposed that the $f$--$\Lambda_2$ relations of strange stars and neutron stars differ irrespective of the EoS, so that a simultaneous measurement of $f_2$ and $\Lambda_2$ in a binary inspiral could provide smoking-gun evidence for strange stars~\cite{pradhan2023}. Gao \textit{et al.}, in contrast, found that the neutron-star universal relations also hold for quark stars and that the resulting difference in dynamical-tide dephasing is undetectable even with next-generation detectors~\cite{gao}. Related degeneracies are known for whole EoS families: different barotropes can share nearly identical tidal curves while differing in radius~\cite{raithel}, twin-star branches can be compared at equal mass~\cite{twins}, and the residuals of frequency relations depend on the density profile and on the slope of the mass--radius curve~\cite{residual,chu}. An entire mass--radius sequence determines a barotrope~\cite{lindblom92}, but a single star is described by only a few numbers.

Here we ask a sharp version of the question: can the mass, $\Lambda_2$ and $f_2$ of one star decide whether it is self-bound? We answer it by construction. Starting from the calibrated analytic nuclear barotrope of Chen, Zhou and Zhang (CZZ)~\cite{czz}, we tune a strange-quark-matter EoS, realized inside the bulk stability window, until the two stars share all three quantities in full general relativity (Sec.~\ref{sec:coincidence}). We then quantify how robust the coincidence is to the strange-quark mass, a realistic crust, the self-bound stiffness and non-barotropic perturbations, separating numerical from physical uncertainties (Sec.~\ref{sec:budget}); measure how much $f_2$ adds to $(M,\Lambda_2)$ across EoS shapes, including when the radius is also fixed (Sec.~\ref{sec:information}); follow both EoSs into binaries and compute noise-weighted mismatches (Sec.~\ref{sec:binary}); and identify the observables that do separate the two stars (Sec.~\ref{sec:breaking}). Section~\ref{sec:methods} defines the models and methods. Numerical validation, a comparison across EoS shapes and a real-frequency scattering check are collected in the appendices.

%=====================================================================
\section{Models and methods}\label{sec:methods}
%=====================================================================
We use $G=c=1$ except where physical units are quoted. Hatted energy densities and pressures are in units of $r_\odot^{-2}$, with $r_\odot=G\ms/c^2=1.47662504$~km. Unless stated otherwise, stars are cold, nonrotating and composed of a single barotropic perfect fluid.

\subsection{Nuclear benchmark and its surface}\label{sec:czz}
The CZZ family~\cite{czz} is the smooth two-power barotrope
\begin{equation}
 \hat\epsilon=a_1\hat p^{\,g_1}+a_2\hat p^{\,g_2},
 \label{eq:czz}
\end{equation}
calibrated against nuclear-physics and astrophysical constraints without specifying a microscopic composition. We adopt its maximum-a-posteriori (MAP) coefficients (Table~\ref{tab:czz}) and keep them fixed for all material comparisons. The median and mean representatives enter only the shape comparison of Appendix~\ref{app:shapes}. The MAP sequence reaches $M_{\max}=2.444\,\ms$. At $1.4\,\ms$ it has $R=11.329$~km, $\Lambda_2=516.9$ and a full-GR $f_2=1.7165$~kHz.

\begin{table}[tbp]
\caption{CZZ coefficients of Eq.~\eqref{eq:czz}~\cite{czz}.}\label{tab:czz}
\setlength{\tabcolsep}{4pt}
\centering\tablebodyfont
\begin{tabular}{@{}lrrrr@{}}
\toprule
Shape & $a_1$ & $g_1$ & $a_2$ & $g_2$\\\midrule
MAP & 0.00532781&0.191371&84.4709&1.61011\\
Median &0.00783&0.22713&3743.13781&2.20319\\
Mean &0.00786&0.22835&1958.45576&2.09076\\
\botrule
\end{tabular}
\end{table}

At low pressure $\epsilon\propto p^{g_1}$ with $g_1\simeq0.19$, an effective polytropic index $n=g_1/(1-g_1)\simeq0.24$. When Eq.~\eqref{eq:czz} is continued to zero pressure, as in Ref.~\cite{czz}, the density therefore stays high until very close to the surface. Compared with the unified SLy EoS~\cite{dh,hp}, the MAP continuation is denser by factors of 15, $2\times10^3$ and $3\times10^6$ at $P=10^{31}$, $10^{28}$ and $10^{22}$~dyn\,cm$^{-2}$. In the $1.4\,\ms$ MAP star, matter below $1.3\times10^{14}$~g\,cm$^{-3}$ forms a layer only 11~m thick that contains $9\times10^{-4}\,\ms$. For SLy the same layer is 0.92~km thick and contains $1.7\times10^{-2}\,\ms$. The analytic MAP star has a zero surface density, but its outer structure is nearly that of a self-bound star.

To obtain neutron stars with a realistic crust we therefore also attach the SLy fluid envelope~\cite{hp}. For a matching density $\rho_h=10^{13}$ or $10^{14}$~g\,cm$^{-3}$, let $p_0$ and $p_1$ be the SLy pressures at $\rho_h/10$ and $\rho_h$. For $p_0<p<p_1$ we use
\begin{equation}
 \epsilon(p)=\epsilon_{\rm SLy}(p)\left[\frac{\epsilon_{\rm MAP}(p)}{\epsilon_{\rm SLy}(p)}\right]^{3t^2-2t^3},
 \label{eq:envelope}
\end{equation}
with $t=\ln(p/p_0)/\ln(p_1/p_0)$, SLy below $p_0$ and MAP above $p_1$. Both $\epsilon$ and $\dd\epsilon/\dd p$ are continuous, and the squared sound speed in the transition is positive and below 0.01. We call these models crust-completed MAP stars. The envelopes are fluid; elastic stresses are not included.

\subsection{Self-bound quark matter}\label{sec:quark}
A bare strange star with massless quarks obeys the linear barotrope $\epsilon=3p+4B$~\cite{afo,hzs,bag}. We realize it with the phenomenological grand potential~\cite{alford2005,auditweiss}
\begin{equation}
 p=\frac{a_4}{4\pi^2}\sum_{i=u,d,s}\mu_i^4+\frac{\mu_e^4}{12\pi^2}-B
 \label{eq:masslessbag}
\end{equation}
in units $\hbar=c=1$, with $B$ in MeV$^4$; a tabulated $B$ in MeV\,fm$^{-3}$ is multiplied by $(\hbar c)^3$. The pressure above the vacuum term is homogeneous of degree four in the chemical potentials, so $\epsilon=\sum_i\mu_i\partial p/\partial\mu_i-p=3p+4B$ for any constant $a_4$. At fixed $B$ and central pressure the gravitational observables are therefore independent of $a_4$, whereas bulk stability is not.

Following the conventional criteria~\cite{farhi,auditweiss} we require neutral three-flavor matter at zero pressure to have $E/A\le930$~MeV and neutral two-flavor matter to have $E/A\ge934$~MeV. At $p=0$, $E/A=\mu_B$. For massless three-flavor matter $\mu_e=0$ and
\begin{equation}
 \mu_B^{uds}=3\left(\frac{4\pi^2B}{3a_4}\right)^{1/4}.
\end{equation}
For two flavors, with $x=\mu_e/\mu_u$, weak equilibrium $\mu_d=\mu_u+\mu_e$ and neutrality $a_4[2-(1+x)^3]=x^3$,
\begin{equation}
 \mu_B^{ud}=(3+2x)\left(\frac{4\pi^2B}{a_4[1+(1+x)^4]+x^4/3}\right)^{1/4}.
\end{equation}

A finite strange-quark mass changes the barotrope itself. We take $m_u=m_d=0$, $m_s=95$~MeV, free massless electrons, $\mu_d=\mu_s\equiv\mu$, $\mu_u=\mu-\mu_e$ and $k=(\mu^2-m_s^2)^{1/2}$. Applying the same factor $a_4$ to the strange-quark term as a phenomenological choice,
\begin{align}
 p+B&=a_4\left[\frac{\mu_u^4+\mu^4}{4\pi^2}+P_s\right]+\frac{\mu_e^4}{12\pi^2},\\
 P_s&=\frac{\mu k(2\mu^2-5m_s^2)+3m_s^4\ln[(\mu+k)/m_s]}{8\pi^2}.
\end{align}
Neutrality gives $a_4(2\mu_u^3-\mu^3-k^3)=\mu_e^3$, so that $n_B=a_4(\mu_u^3+\mu^3+k^3)/(3\pi^2)$ and $\epsilon=(3\mu-\mu_e)n_B-p$. Perturbations follow this neutral equilibrium branch.

\subsection{Scale map and matching}\label{sec:scale}
For any reference EoS $\epsilon_0(p)$ and positive constant $s$, the map
\begin{equation}
 \begin{aligned}
 \epsilon_s(p)&=s\,\epsilon_0(p/s),\\
 (r,m)&\mapsto\frac{(r,m)}{\sqrt s},\qquad
 (p,\epsilon)\mapsto s(p,\epsilon)
 \end{aligned}
 \label{eq:scale}
\end{equation}
leaves the Tolman--Oppenheimer--Volkoff equations invariant. At equal scaled central pressure it preserves the compactness $C=M/R$, the Love numbers $k_\ell$, $\Lambda_\ell$ and the complex $M\omega_\ell$, so $f_\ell\mapsto\sqrt s\,f_\ell$ and $\tau_\ell\mapsto\tau_\ell/\sqrt s$. For the bag EoS $s=B/B_0$; for CZZ, $a_i\mapsto a_is^{1-g_i}$. To place a target $(M_*,\Lambda_*)$ we locate $\Lambda_2(p_c)=\Lambda_*$ on the increasing-mass branch and set $s=(M_0/M_*)^2$, where $M_0$ is the mass found there. Rescaling changes dimensional EoS parameters; results at published normalizations are listed separately in Appendix~\ref{app:shapes}.

\subsection{Static tides}\label{sec:tides}
For the even-parity quadrupolar tide $y=rH'/H$ obeys $ry'+y^2+Fy+r^2Q=0$ with $y(0)=2$, where
\begin{align}
 F&=\frac{1+4\pi r^2(p-\epsilon)}{1-2m/r},\\
 Q&=\frac{4\pi[5\epsilon+9p+(\epsilon+p)\,\dd\epsilon/\dd p]-6/r^2}{1-2m/r}-4(\Phi')^2,
\end{align}
and $e^{\Phi(R)}=\sqrt{1-2C}$. Matching to vacuum gives $k_2$ and $\Lambda_2=2k_2/(3C^5)$~\cite{hinderer}. At the surface of a self-bound star $y(R^+)=y(R^-)-4\pi R^3\epsilon_s/M$~\cite{postnikov,dn}. A related correction is needed when the density vanishes slowly. In the variable $\ln p$ the compressibility term of $\dd y/\dd\ln p$ is $4\pi r^3p(\dd\epsilon/\dd p)/(m+4\pi r^3p)$, and integrating the omitted outer layer gives
\begin{equation}
 y(p\to0)-y(p_{\min})\simeq-\frac{4\pi R^3\epsilon(p_{\min})}{M}.
 \label{eq:tail}
\end{equation}
For the $1.4\,\ms$ MAP star, surface cutoffs $x\equiv-\ln(p_{\min}/p_c)=20$, 40, 60 and 90 give $\Lambda_2=489.45$, 516.26, 516.86 and 516.88 without this term, whereas adding Eq.~\eqref{eq:tail} at $x=20$ already gives 516.877. The value 512.1 quoted in Ref.~\cite{czz} is reproduced to 0.1\% by truncating at $x\simeq30$ (512.74), consistent with a truncated tail. We recompute every static quantity from the barotrope and use 512.1 only as a common reference target in Sec.~\ref{sec:information} and Appendix~\ref{app:shapes}.

\subsection{Radiating modes}\label{sec:modes}
We define $\omega_\ell=2\pi f_\ell+i/\tau_\ell$ with time dependence $e^{i\omega t}$, so $\tau_\ell$ is the amplitude damping time. Fluid displacements and polar metric perturbations are solved together in the Lindblom--Detweiler formulation~\cite{ld,dl}. Interior solutions that are regular at the center and satisfy $\Delta p=0$ at the surface are matched to the vacuum Zerilli equation, and a quasinormal mode is a complex root of the incoming-wave amplitude with outgoing behavior $e^{-i\omega r_*}$ at infinity. In the variables of Ref.~\cite{dl}, with $r^\ell$ factored out, the pressure variable is
\begin{equation}
 X=\omega^2(\epsilon+p)e^{-\Phi}V-\frac{p'}{r}e^{\Phi-\lambda/2}W+\tfrac12(\epsilon+p)e^{\Phi}H_0 ,
\end{equation}
with $e^{-\lambda}=1-2m/r$. The $\omega^2V$ term follows from the linearized Euler equation.\footnote{Earlier versions of this e-print misprinted the power of $\omega$ in this relation.} Implementation and validation are described in Appendix~\ref{app:numerics}.

As a control we also solve the relativistic Cowling problem~\cite{cowling}, in which metric perturbations are frozen:
\begin{align}
 W'&=\frac{\dd\epsilon}{\dd p}\left[\omega^2r^2 e^{\lambda/2-2\Phi}V+\Phi'W\right]\nonumber\\
   &\quad-\ell(\ell+1)e^{\lambda/2}V,\\
 V'&=2\Phi'V-e^{\lambda/2}W/r^2 ,
\end{align}
with $W\propto r^{\ell+1}$, $V\propto-r^\ell/\ell$ at the center and a vanishing bracket at the surface. Cowling frequencies are real; we use them only to show how the approximation distorts small frequency differences.

%=====================================================================
\section{A quark star and a neutron star with identical \texorpdfstring{$(M,\Lambda_2,f_2)$}{(M, Lambda2, f2)}}\label{sec:coincidence}
%=====================================================================
\subsection{Locating the coincidence}\label{sec:cross}
\begin{figure*}[tp]
\centering\includegraphics[width=\textwidth]{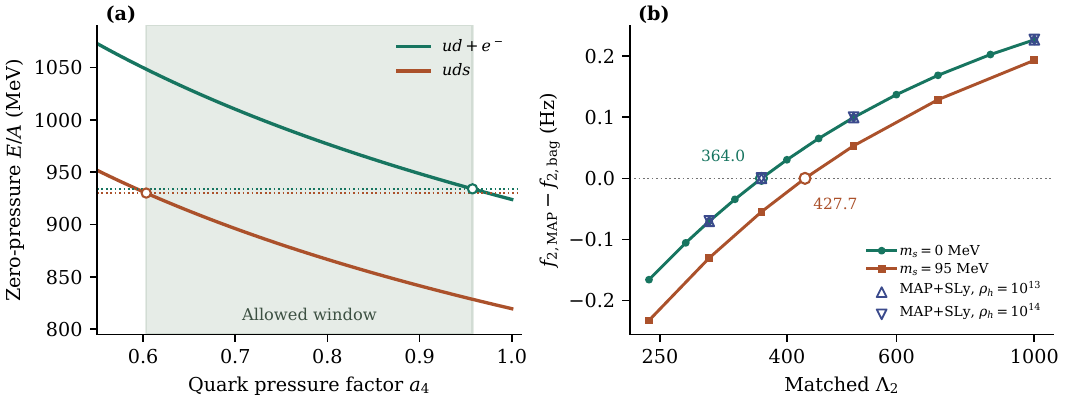}
\caption{Bulk stability and quadrupolar coincidences. (a)~Zero-pressure energy per baryon of neutral two-flavor ($ud+e^-$) and three-flavor ($uds$) matter at the massless coincidence, $B=55.107$~MeV\,fm$^{-3}$. Both criteria, $E_{ud}/A>934$~MeV and $E_{uds}/A<930$~MeV, hold in the shaded interval $0.603<a_4<0.957$. (b)~Full-GR frequency difference between MAP and bag stars of equal mass and $\Lambda_2$, at the physical mass of the MAP star. For each target $\Lambda_2$ the bag constant is adjusted to that mass. Open circles mark the refined roots for $m_s=0$ and 95~MeV ($a_4=0.90$). Triangles show the crust-completed MAP stars of Eq.~\eqref{eq:envelope} with the massless bag. Lines guide the eye.}
\label{fig:crossing}
\end{figure*}

Because $M\omega_2$ and $\Lambda_2$ are invariant under Eq.~\eqref{eq:scale}, the comparison between MAP and the linear bag EoS reduces to one dimensionless function. We denote by $\Delta f_2(\Lambda_2)$ the MAP-minus-bag difference of the real frequencies of two stars with the same $M$ and $\Lambda_2$. At a reference mass of $1.4\,\ms$ it equals $-0.078$~Hz at $\Lambda_2=300$, $+0.100$~Hz at 512.1 and $+0.199$~Hz at 1000. Figure~\ref{fig:crossing}(b) shows $\Delta f_2$ at the physical mass of the MAP star: for $250\lesssim\Lambda_2\le1000$ its magnitude stays below 0.25~Hz.

Its zero lies at $\Lambda_*=364.035$, where the MAP star has $M=1.4957\,\ms$. The choice $B=55.107$~MeV\,fm$^{-3}$ ($B^{1/4}=143.4$~MeV) places the bag star at the same mass. The two stars then share $M$, $\Lambda_2$ and ${\rm Re}\,f_2=1732.87$~Hz, although one of them has a density jump of $4B$ at its surface. Their damping times differ by a fraction $6\times10^{-5}$ (0.22317 and 0.22316~s) and their radii by 0.20~km (11.49 and 11.29~km). The first row of Table~\ref{tab:coincidence} summarizes the coincidence. When this pair is followed to other masses, $B$ and the MAP coefficients are held fixed.

\subsection{Microphysical realization}\label{sec:micro}
For a free massless gas ($a_4=1$) the tuned $B$ gives $E/A_{ud}=923.9$~MeV, below the two-flavor bound, so ordinary nuclei would be unstable. The same $B$ satisfies both bulk criteria for $0.603\lesssim a_4\lesssim0.957$ [Fig.~\ref{fig:crossing}(a)]. We adopt $a_4=0.90$, for which $(E/A_{ud},E/A_{uds})=(948.6,841.5)$~MeV; the value of $a_4$ does not affect any gravitational observable. The bag sequence reaches $M_{\max}=2.049\,\ms$, consistent within one standard deviation with the $2.08\pm0.07\,\ms$ of PSR~J0740+6620~\cite{fonseca}. The bulk criteria constrain equilibrium energies only; nucleation and conversion rates require additional microphysics.

\subsection{Strange-quark mass and crust}\label{sec:variants}
For $m_s=95$~MeV and $a_4=0.90$ the squared sound speed of the matched quark star varies from 0.3214 at the surface to 0.3248 at the center. Because $B$ now also changes $m_s/B^{1/4}$, we solve directly for the physical $B$ and central pressure that reproduce the MAP mass and $\Lambda_2$ at each trial deformability. The frequency difference changes sign between $\Lambda_2=364.035$ ($-0.055$~Hz) and 512.1 ($+0.053$~Hz), and its root gives a second coincidence at $M=1.4513\,\ms$, $\Lambda_2=427.73$ and $f_2=1725.30$~Hz, with $B=52.147$~MeV\,fm$^{-3}$ (Table~\ref{tab:coincidence}). Its zero-pressure energies per baryon are 935.6~MeV for two flavors and 851.5~MeV for three. The two-flavor criterion is met by only 1.6~MeV, so this example lies at the edge of the stability window.

Replacing the low-density MAP continuation by the SLy envelope moves the coincidence only to $\Lambda_*=364.02$ ($\rho_h=10^{13}$~g\,cm$^{-3}$) and 363.41 ($10^{14}$~g\,cm$^{-3}$). The matched mass, $B$ and $f_2$ change by less than $6\times10^{-4}\,\ms$, $7\times10^{-3}$~MeV\,fm$^{-3}$ and 0.01~Hz. The neutron-star radius, however, grows to 11.92 and 12.11~km, so that the coincident stars now differ in radius by 0.63 and 0.82~km. At all four sampled deformabilities the crust-completed values of $\Delta f_2$ lie on the analytic-surface curve [Fig.~\ref{fig:crossing}(b)]; at $\Lambda_2=512.1$ they differ from it by less than $10^{-3}$~Hz. The crust therefore changes the radius at fixed $(M,\Lambda_2)$ by several hundred meters while leaving $f_2$ essentially unchanged.

\begin{table*}[tp]
\caption{Self-bound stars matched to the MAP sequence. Each row gives a common mass $M$, tidal deformability $\Lambda_2$ and real quadrupolar frequency $f_2$. $\Delta R=R_{\rm NS}-R_{\rm QS}$ and $\Delta f_3=f_{3,\rm QS}-f_{3,\rm NS}$; $M_{\max}$ refers to the quark-star sequence and $\rho_h$ is in g\,cm$^{-3}$. The digits record numerical solutions of the specified models; Table~\ref{tab:budget} gives their physical uncertainty.}\label{tab:coincidence}
\setlength{\tabcolsep}{4pt}
\centering\tablebodyfont
\begin{tabular}{@{}lrrrrrrr@{}}
\toprule
Model & $B$ (MeV\,fm$^{-3}$) & $M/\ms$ & $\Lambda_2$ & $f_2$ (Hz) & $\Delta R$ (km) & $\Delta f_3$ (Hz) & $M_{\max}/\ms$\\\midrule
Analytic surface; $m_s=0$ & 55.107 & 1.4957 & 364.04 & 1732.87 & 0.202 & 1.88 & 2.049\\
SLy envelope, $\rho_h=10^{13}$; $m_s=0$ & 55.107 & 1.4957 & 364.02 & 1732.87 & 0.634 & 1.88 & 2.049\\
SLy envelope, $\rho_h=10^{14}$; $m_s=0$ & 55.101 & 1.4963 & 363.41 & 1732.88 & 0.823 & 1.93 & 2.049\\
Analytic surface; $m_s=95$~MeV & 52.147 & 1.4513 & 427.73 & 1725.30 & 0.201 & 1.87 & 2.025\\
\botrule
\end{tabular}
\end{table*}

\subsection{Error budget}\label{sec:budget}
Table~\ref{tab:budget} separates numerical from physical uncertainties. Numerical refinement changes $\Lambda_*$ by less than $10^{-5}$ and $f_2$ by less than $10^{-5}$~Hz (Appendix~\ref{app:numerics}). Physical changes move the coincidence much further, because the slope $\partial\Delta f_2/\partial\Lambda_2\simeq9.8\times10^{-4}$~Hz is so small. Changing the squared sound speed of the self-bound fluid, $p=c_s^2(\epsilon-\epsilon_0)$, by $\pm0.005$ around $1/3$ moves $\Lambda_*$ to 336 and 394. The strange-quark mass moves it to 428.

Non-barotropic perturbations matter much less. In neutron-star matter the frozen-composition and equilibrium sound speeds differ by $c^2_{\rm ad}-c^2_{\rm eq}\lesssim0.01$~\cite{zhao2025}; see also Ref.~\cite{montefusco}. We keep the MAP equilibrium and static tide but use the adiabatic index $\Gamma_1=(1+\kappa)\Gamma_{\rm eq}$ in its perturbation equations. For $\kappa=0.02$ and 0.05 this raises $f_2$ by only 0.0018 and 0.0045~Hz, because the $f$ mode is nearly divergence free~\cite{zhao2025}, and the coincidence moves to $\Lambda_*\simeq362$ and 359. The quark star needs no such correction: the composition of massless three-flavor matter is fixed.

Finally, an inspiral probes the conservative tidal response rather than the quasinormal mode. Gao and Steinhoff found a 5.4~Hz offset between the two for one hybrid star~\cite{gaost}, more than an order of magnitude larger than $|\Delta f_2|$; we have not computed this offset for the present stars. Rotation, elasticity and superfluidity are also neglected. The existence of a coincidence is therefore robust, but its location depends on the model at the 10--20\% level in $\Lambda_2$.

\begin{table*}[tp]
\caption{Error budget for the massless coincidence. $\Lambda_*$ and $M_*$ locate the coincidence; the last column lists other changes.}\label{tab:budget}
\centering\tablebodyfont
\begin{tabular}{@{}>{\raggedright\arraybackslash}p{0.35\textwidth}ll>{\raggedright\arraybackslash}p{0.25\textwidth}@{}}
\toprule
Change & $\Lambda_*$ & $M_*/\ms$ & Remark\\\midrule
Reference: analytic surface, $m_s=0$, $c_s^2=1/3$ & 364.035 & 1.4957 & $f_2=1732.87$~Hz\\
Numerical refinement (interior, exterior, surface cutoff) & $|\delta\Lambda_*|<10^{-5}$ & $<10^{-6}$ & $|\delta f_2|<10^{-5}$~Hz\\
SLy envelope, $\rho_h=10^{13}$ / $10^{14}$~g\,cm$^{-3}$ & 364.02 / 363.41 & 1.4957 / 1.4963 & $R_{\rm NS}$ larger by 0.43 / 0.62~km\\
Strange-quark mass $m_s=95$~MeV & 427.73 & 1.4513 & $B=52.15$~MeV\,fm$^{-3}$\\
Self-bound $c_s^2=1/3-0.005$ / $1/3+0.005$ & 394.46 / 336.22 & 1.4735 / 1.5179 & $M_{\max}=2.035$ / $2.063\,\ms$\\
Neutron-star perturbations with $\Gamma_1=(1+\kappa)\Gamma_{\rm eq}$, $\kappa=0.02$ / 0.05 & 362.04 / 359.19 & 1.4972 / 1.4994 & $\delta f_2=+0.0018$ / $+0.0045$~Hz\\
Conservative tidal response instead of quasinormal mode & not computed & & offset of 5.4~Hz for one hybrid star~\cite{gaost}\\
\botrule
\end{tabular}
\end{table*}

%=====================================================================
\section{\texorpdfstring{How much does $f_2$ add to $(M,\Lambda_2)$?}{How much does f2 add to (M, Lambda2)?}}\label{sec:information}
%=====================================================================
The coincidence does not rely on fine tuning: $\Delta f_2$ is small everywhere. Across $250\lesssim\Lambda_2\le1000$ the MAP and bag frequencies at equal $(M,\Lambda_2)$ differ by less than 0.25~Hz, about $1.5\times10^{-4}$ of $f_2$, and the crust-completed MAP stars follow the same curve. The intrinsic linewidth of the mode, $1/(\pi\tau_2)\simeq1.3$~Hz, is already larger, and networks of future detectors are expected to measure $f$-mode frequencies to within tens of hertz~\cite{binpratten}. A simultaneous measurement of $\Lambda_2$ and $f_2$ therefore cannot separate these two stars at any mass in this range.

To test whether this is specific to the pair, Appendix~\ref{app:shapes} compares seven EoS shapes, and SLy, rescaled to the same $M=1.4\,\ms$ and $\Lambda_2=512.1$. The radii of the seven shapes span 2.85~km and their $f_2$ span 18.5~Hz. The spread is dominated by shapes whose rescaled sequences do not reach $2\,\ms$: the quartic scalar fluid, the WSS fit and the $\Gamma=2$ polytrope. The five models that do---the bag, the three CZZ representatives and SLy---agree within 0.54~Hz. In the Cowling approximation the three CZZ shapes alone span 15~Hz instead of 0.15~Hz, so frozen-metric calculations exaggerate the EoS sensitivity by two orders of magnitude here.

Fixing the radius as well does not change this conclusion. Within the two-power family we hold $g_2=1.61011$, vary $g_1$ around the MAP value, and solve for $a_2$ and $p_c$ such that $(M,R,\Lambda_2)=(1.4\,\ms,\,11.313~{\rm km},\,512.1)$; the scale map then restores the mass. With $a_1\hat p^{g_1}$ held fixed at $\hat p=1.4894816\times10^{-4}$, the Jacobian of $(\ln C,\ln\Lambda_2)$ with respect to $(\ln a_2,\ln p_c)$ at the reference model has determinant 0.0101, stable under a two-decade change of the finite-difference step, so the implicit-function theorem leaves one free exponent. For $0<g_1<1<g_2$ the global maximum of $c_s^2$ lies at $\hat p=[a_1g_1(1-g_1)/(a_2g_2(g_2-1))]^{1/(g_2-g_1)}$, which gives a grid-independent causality check.

\begin{figure*}[tp]
\centering\includegraphics[width=\textwidth]{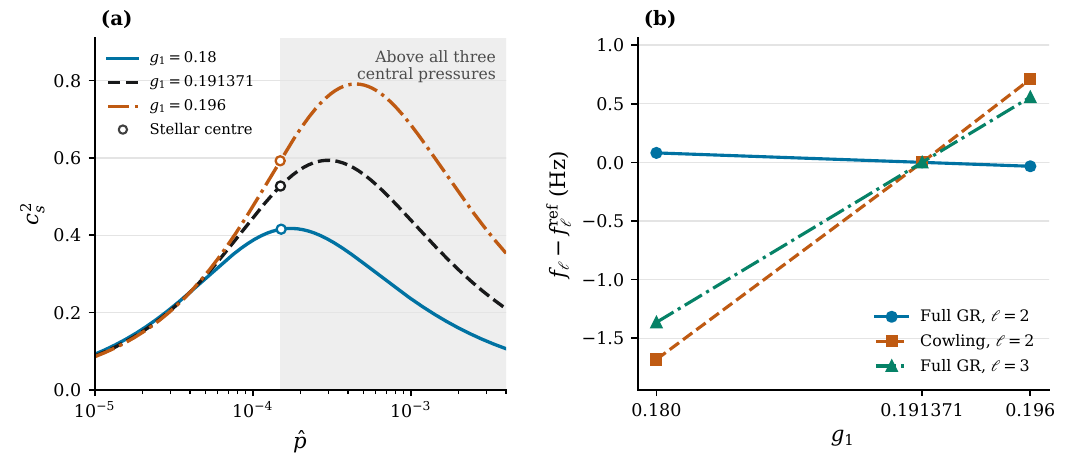}
\caption{Three smooth, causal EoSs with the same $M$, $R$ and $\Lambda_2$. (a)~Squared sound speed; open circles mark the matched central pressures and the shaded range lies above all three centers. (b)~Frequency shifts from the middle model, computed on the matched backgrounds.}\label{fig:triple}
\end{figure*}

Table~\ref{tab:triple} lists three such EoSs. Between the endpoints the maximum $c_s^2$ changes from 0.42 to 0.79 and $M_{\max}$ from 2.14 to $2.65\,\ms$, yet $f_2$ changes by only $-0.114$~Hz in full GR. The Cowling approximation predicts $+2.39$~Hz, with the opposite sign and 21 times the magnitude. The octupolar frequency changes by $+1.92$~Hz. Most of the change in stiffness lies at pressures above the matched centers (Fig.~\ref{fig:triple}).

\begin{table*}[tp]
\caption{Smooth two-power EoSs with the same $M=1.4\,\ms$, $R=11.313$~km and $\Lambda_2=512.1$. The maximum $c_s^2$ is over the complete barotrope; $M_{\max}$ belongs to each sequence. Frequencies and damping times are full-GR values for the matched stars.}\label{tab:triple}
\centering\tablebodyfont
\begin{tabular}{@{}rrrrrrr@{}}
\toprule
$g_1$ & $\max c_s^2$ & $M_{\max}/\ms$ & $f_2$ (Hz) & $f_3$ (Hz) & $\tau_2$ (s) & $\tau_3$ (s)\\\midrule
0.18 & 0.418 & 2.137 & 1720.06 & 2357.45 & 0.2412 & 19.96\\
0.191371 & 0.594 & 2.440 & 1719.98 & 2358.81 & 0.2412 & 19.94\\
0.196 & 0.791 & 2.652 & 1719.95 & 2359.37 & 0.2413 & 19.93\\
\botrule
\end{tabular}
\end{table*}

These results are incompatible with an EoS-independent separation of strange stars and neutron stars by the $f$--$\Lambda_2$ relation~\cite{pradhan2023}: a bag model inside the bulk stability window and a calibrated nuclear barotrope share the relation within 0.25~Hz over the inspiral-relevant range. Where such differences appear in other model sets, they reflect particular EoS choices rather than self-boundness itself. Matching like multipoles also matters. At fixed $\Lambda_3$ instead of $\Lambda_2$ the $f_3$ spread of the seven shapes falls from 186 to 63~Hz while the $f_2$ spread rises to 86~Hz (Appendix~\ref{app:shapes}), consistent with the diagonal $f$--Love relations of Ref.~\cite{chan}.

%=====================================================================
\section{Binary inspirals}\label{sec:binary}
%=====================================================================
\subsection{Equal leading tidal parameter}\label{sec:equal}
\begin{figure*}[tp]
\centering\includegraphics[width=\textwidth]{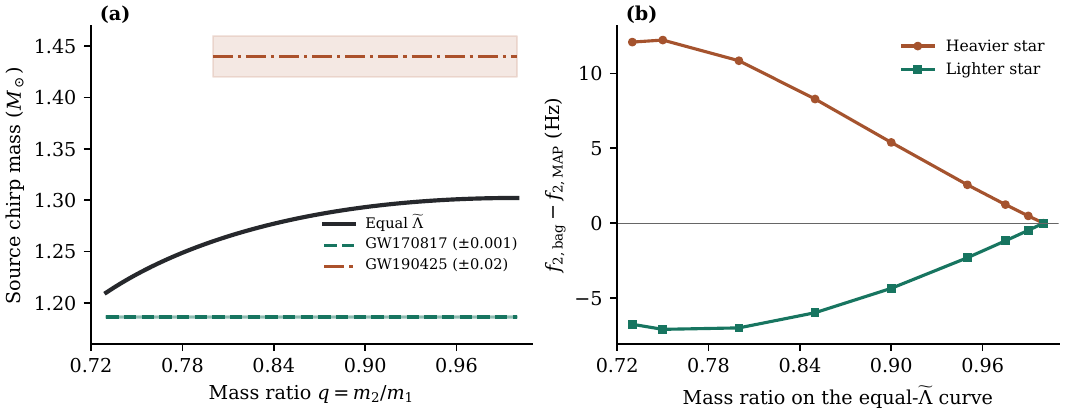}
\caption{Binaries governed by the fixed bag and MAP EoSs. (a)~Curve on which their leading tidal parameters agree. Horizontal bands show the published marginal source chirp-mass intervals over the low-spin mass-ratio ranges of GW170817 and GW190425~\cite{binevent17,binevent19}; the GW170817 interval is narrower than its line. (b)~Differences of the component source-frame frequencies along the curve.}
\label{fig:binary}
\end{figure*}

Consider two nonspinning quasicircular binaries with identical component masses $M_A\ge M_B$, one built from MAP and the other from the tuned bag EoS, both held fixed. Let $q=M_B/M_A$, $M_T=M_A+M_B$, $X_i=M_i/M_T$, $\eta=X_AX_B$ and $\mathcal M=M_T\eta^{3/5}$. Throughout this section $\Delta$ denotes bag minus MAP. The leading tidal phase depends on~\cite{flanagan,binwade}
\begin{equation}
 \Ltil=\frac{16}{13}\sum_{i\ne j}(X_i+12X_j)X_i^4\Lambda_i ,
\end{equation}
which can coincide for different barotropes through cancellations between unequal components~\cite{raithelq,yagibinary}. With $\delta=(1-q)/(1+q)$ and $\bar M=2^{1/5}\mathcal M$, the component masses at fixed $\mathcal M$ are $M_{A,B}=\bar M(1\pm\delta)(1-\delta^2)^{-3/5}$, and
\begin{align}
 \Ltil=\Lambda_2(\bar M)+\Big[&\tfrac{34}{13}\Lambda_2(\bar M)+\tfrac{244}{65}\bar M\Lambda_2'(\bar M)\nonumber\\
 &+\tfrac12\bar M^2\Lambda_2''(\bar M)\Big]\delta^2+O(\delta^4),
 \label{eq:binary_expansion}
\end{align}
where primes denote derivatives along a fixed EoS sequence and odd terms cancel by exchange symmetry. At the coincidence $\Delta\Lambda_2'=-61.9\,\ms^{-1}$ and the brackets differ by $-226.6$, so the curve of equal $\Ltil$ is locally $\mathcal M_{\rm eq}/\ms=1.3021-3.19\,\delta^2+O(\delta^4)$. Figure~\ref{fig:binary}(a) shows the unexpanded solution. At $q=0.8$ it lies at $\mathcal M=1.2598\,\ms$ with $\Ltil=424.2$; it reaches the GW170817 chirp mass at $q\simeq0.71$, just below the low-spin interval. Along the curve the component frequencies differ at first order in $\delta$, by up to $+12$~Hz for the heavier star and $-7$~Hz for the lighter one [Fig.~\ref{fig:binary}(b)].

\subsection{Residual dephasing}\label{sec:dephasing}
On this curve the leading term $\Psi_5=-117\Ltil v^5/(256\eta)$ of the Fourier phase is identical for both binaries, with $v=[\pi(1+z)M_Tf]^{1/3}$. The remaining differences come from the next tidal order $\Psi_6$, for which we use Eqs.~(4)--(6) of Ref.~\cite{binwade}, and from the quadrupolar $f$-mode response~\cite{hinderer2016,steinhoff2016}. For the latter we use the frequency-domain model of Schmidt and Hinderer~\cite{binfmtidal},
\begin{align}
 \Psi_{f_2}=&-\frac{10\sqrt3\,\pi-27-30\ln2}{96\eta}\,v^{11}\nonumber\\
 &\times\sum_{i}\frac{\Lambda_iX_i^6(155-147X_i)}{\bar\omega_i^{\,2}},
 \label{eq:binary_phase}
\end{align}
with $\bar\omega_i=2\pi M_if_{2,i}$. Both residuals start at order $\delta^2$ on the curve.

Phase differences at a single frequency depend on the time and phase conventions, so we compare the signals through their overlap maximized over coalescence time and phase,
\begin{equation}
 \begin{aligned}
 \mathcal O&=\max_{t_c}\frac{\left|\int w(f)\,e^{i[\Delta\Psi(f)-2\pi ft_c]}\dd f\right|}{\int w(f)\,\dd f},\\
 w&=\frac{f^{-7/3}}{S_n(f)} .
 \end{aligned}
\end{equation}
The two signals differ by one standard deviation when the signal-to-noise ratio (SNR) reaches $\rho_{1\sigma}=[2(1-\mathcal O)]^{-1/2}$~\cite{lindblom2008}. Only the shape of $S_n$ enters. We use the analytic Advanced LIGO design shape from 10~Hz~\cite{keppel} and, as a bracket that weights high frequencies more strongly, white noise from 20~Hz, both up to 1~kHz, near the validity limit of Eq.~\eqref{eq:binary_phase}~\cite{binfmtidal}. Table~\ref{tab:mismatch} shows that $\rho_{1\sigma}\gtrsim4\times10^4$ everywhere on the curve with $q\ge0.73$. Because this criterion ignores correlations with masses and spins, which can only raise the required SNR, the component frequency differences of several hertz are unobservable in the inspiral, consistent with the quark-star analysis of Ref.~\cite{gao}. Third-generation prospects for measuring the dynamical tides themselves are discussed in Ref.~\cite{bingw3g}.

\begin{table}[tbp]
\caption{Distinguishability on the equal-$\Ltil$ curve. $\rho_{1\sigma}=[2(1-\mathcal O)]^{-1/2}$ is a lower bound on the SNR needed to separate the bag and MAP binaries, for the Advanced LIGO design shape and for white noise, with $f\le1$~kHz.}\label{tab:mismatch}
\setlength{\tabcolsep}{4pt}
\centering\tablebodyfont
\begin{tabular}{@{}rrrrr@{}}
\toprule
$q$ & $\mathcal M/\ms$ & $\Ltil$ & $\rho_{1\sigma}$ (aLIGO) & $\rho_{1\sigma}$ (white)\\\midrule
0.95 & 1.3000 & 366.8 & $2.1\times10^6$ & $1.1\times10^6$\\
0.90 & 1.2931 & 375.9 & $5.0\times10^5$ & $2.7\times10^5$\\
0.80 & 1.2598 & 424.2 & $1.2\times10^5$ & $6.8\times10^4$\\
0.73 & 1.2099 & 511.6 & $7.7\times10^4$ & $4.2\times10^4$\\
\botrule
\end{tabular}
\end{table}

\subsection{Observed chirp masses}\label{sec:events}
At the central source chirp masses of GW170817 ($1.186\,\ms$, $0.73\le q\le1$) and GW190425 ($1.44\,\ms$, $0.80\le q\le1$), taken from the published low-spin summaries~\cite{binevent17,binevent19}, the two EoSs give the $\Ltil$ values of Table~\ref{tab:events}. The change in sign of $\Delta\Ltil$ between the two events is a direct consequence of the coincidence at $\mathcal M\simeq1.30\,\ms$. Both models lie inside the published GW170817 90\% interval $70\le\Ltil\le720$ obtained with a flat $\Ltil$ prior~\cite{binevent17} and below the GW190425 low-spin bound $\Ltil<600$~\cite{binevent19}. Since the CZZ calibration already used GW170817 and NICER information~\cite{czz}, this is a consistency check rather than a test. Here $\Ltil$ itself differs, and the optimistic criterion gives $\rho_{1\sigma}\simeq380$--3200 for GW170817-like and 210--390 for GW190425-like sources, compared with the network SNR of 32.4 of GW170817~\cite{gw170817}. The predicted source-frame $f_2$ of the GW170817 components, 1.67--1.76~kHz, lies above the 90\% lower bounds of 1.39~kHz (heavier star) and 1.48~kHz (lighter star) from the dynamical-tide analysis of Ref.~\cite{binpratten}, which depend on its waveform model and priors.

\begin{table*}[tp]
\caption{Conditional predictions at the central source chirp masses over the low-spin mass-ratio ranges. $\Delta\Ltil$ compares the models at the same $q$. The last two columns give the range of $\rho_{1\sigma}$ over the same $q$ interval, at the redshifts $z=0.0099$ and 0.03.}\label{tab:events}
\centering\tablebodyfont
\begin{tabular}{@{}lrrrrrr@{}}
\toprule
Event & $\mathcal M/\ms$ & $\Ltil_{\rm MAP}$ & $\Ltil_{\rm bag}$ & $\Delta\Ltil$ & $\rho_{1\sigma}$ (aLIGO) & $\rho_{1\sigma}$ (white)\\\midrule
GW170817 & 1.186 & 566.8--595.3 & 568.2--604.4 & $+1.41$ to $+9.07$ & 530--3200 & 380--2200\\
GW190425 & 1.440 & 204.2--208.2 & 193.4--199.8 & $-10.72$ to $-8.44$ & 300--390 & 210--280\\
\botrule
\end{tabular}
\end{table*}

Equation~\eqref{eq:binary_phase} assigns each oscillator the full static $\Lambda_i$ and the real part of the quasinormal frequency and neglects spin. In general relativity the dynamical tidal response is not exhausted by mode excitations~\cite{apostol,hegade2026}, and the conservative response pole can differ from the quasinormal frequency by several hertz~\cite{gaost}. Unless these effects differ between the two stars by several hertz, they do not change the order of magnitude of $\rho_{1\sigma}$ on the equal-$\Ltil$ curve, which is set by residual phase differences below $10^{-3}$~rad.

%=====================================================================
\section{What breaks the degeneracy}\label{sec:breaking}
%=====================================================================
Table~\ref{tab:breaking} collects the differences between the coincident stars, and between the two fixed EoSs at other masses, for observables that are not used in the construction.

\emph{Radius.}---The most direct discriminant is $R$: 0.20~km for the analytic surface and 0.63--0.82~km with an SLy envelope. Because the crust changes $R$ much more than $\Lambda_2$ or $f_2$, a radius measurement at this level has to be interpreted together with a model of the outer layers.

\begin{figure*}[tp]
\centering\includegraphics[width=\textwidth]{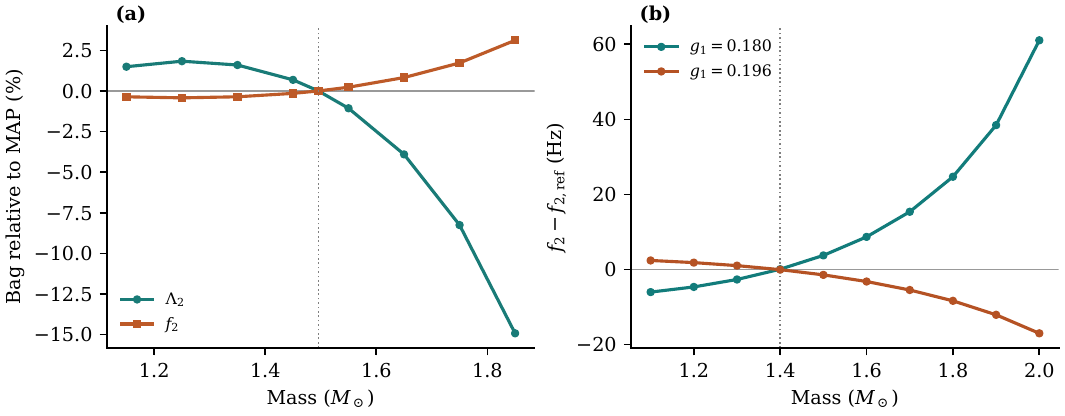}
\caption{Mass dependence with the EoSs held fixed. (a)~Percentage differences of $\Lambda_2$ and $f_2$ between the tuned bag EoS and MAP. (b)~Full-GR $f_2$ of the two endpoint EoSs of Table~\ref{tab:triple} relative to the middle one. Vertical lines mark the matching masses; markers are separate eigenvalue calculations.}\label{fig:sweep}
\end{figure*}

\emph{A second mass.}---With both EoSs fixed [Fig.~\ref{fig:sweep}(a)], $\Lambda_2$ differs by $+1.5\%$ and $f_2$ by $-6$~Hz at $1.15\,\ms$, and by $-15\%$ and $+56$~Hz at $1.85\,\ms$. In the smooth family matched in $(M,R,\Lambda_2)$ at $1.4\,\ms$ [Fig.~\ref{fig:sweep}(b)], the two endpoints separate by 78~Hz at $2\,\ms$, where they have $\Lambda_2=52.5$ and 66.5 and $R=11.90$ and 12.14~km. The mass at which a second measurement is made matters more than sub-hertz precision at one mass.

\emph{Octupole.}---$f_3$ differs by 1.88~Hz (1.93~Hz with the denser envelope). The $\ell=3$ mode decays over $\tau_3\simeq18$~s. For separately excited ringdowns $h\propto e^{-t/\tau+2\pi ift}$ starting at a common time, the phase-maximized overlap in white noise,
\begin{equation}
 \mathcal O=\frac{2\sqrt{\tau_a\tau_b}}{\sqrt{(\tau_a+\tau_b)^2+(2\pi\Delta f\,\tau_a\tau_b)^2}},
\end{equation}
is 0.0093 for the octupoles, whereas $1-\mathcal O=4.3\times10^{-10}$ for the quadrupoles. Exploiting this difference would require a long coherent octupolar signal from an isolated star, whose excitation is not addressed here.

\emph{Moment of inertia.}---As expected from the universality of the $I$--Love relation~\cite{ilove,iloveself}, the coincident stars have $I=1.7576$ and $1.7578\times10^{45}$~g\,cm$^2$, a fractional difference of $1.1\times10^{-4}$ that the envelope leaves unchanged. At the mass of PSR~J0737$-$3039A, $M_A=1.338183\,\ms$, the fixed EoSs give $I_{45}=1.469$ (MAP) and 1.475 (bag), and SLy gives 1.287. The 0.47\% MAP--bag difference corresponds to $2.6\times10^{-6}$~deg\,yr$^{-1}$ in the Lense--Thirring contribution to the periastron advance, using the coefficient of Ref.~\cite{kramer}; all three values lie below the 90\% bound $I_A<3.0\times10^{45}$~g\,cm$^2$~\cite{kramer}.

\emph{Binary inspiral.}---On the equal-$\Ltil$ curve $\rho_{1\sigma}\gtrsim4\times10^4$ (Sec.~\ref{sec:dephasing}).

The coincidence is thus broken by global structure---the radius and the dependence on mass---and not by the $f$ mode at one mass, in line with the result that a mass--radius curve rather than a single star determines the barotrope~\cite{lindblom92}.

\begin{table*}[tp]
\caption{Observables not used in the construction. Differences are bag minus MAP at the massless coincidence unless a mass is given; values in parentheses use the SLy envelopes with $\rho_h=10^{13}$ and $10^{14}$~g\,cm$^{-3}$.}\label{tab:breaking}
\centering\tablebodyfont
\setlength{\tabcolsep}{4pt}
\begin{tabular}{@{}>{\raggedright\arraybackslash}p{0.28\textwidth}>{\raggedright\arraybackslash}p{0.33\textwidth}>{\raggedright\arraybackslash}p{0.31\textwidth}@{}}
\toprule
Observable & Difference & Scale for comparison\\\midrule
Radius at $M_*$ & $-0.202$~km ($-0.634$, $-0.823$~km) & \\
Quadrupolar damping time at $M_*$ & $-1.3\times10^{-5}$~s (fraction $-5.9\times10^{-5}$) & linewidth $1/(\pi\tau_2)\simeq1.3$~Hz\\
Octupolar frequency at $M_*$ & $+1.88$~Hz ($+1.88$, $+1.93$~Hz) & $\tau_3\simeq18$~s; ringdown overlap 0.0093\\
Moment of inertia at $M_*$ & fraction $+1.1\times10^{-4}$ (unchanged) & \\
Moment of inertia at $1.338\,\ms$ & fraction $+4.7\times10^{-3}$ & $I_A<3.0\times10^{45}$~g\,cm$^2$ (90\%)~\cite{kramer}\\
$\Lambda_2$ and $f_2$ at $1.15\,\ms$ & $+1.5\%$, $-6.0$~Hz & $f$-mode precision of tens of hertz~\cite{binpratten}\\
$\Lambda_2$ and $f_2$ at $1.85\,\ms$ & $-15\%$, $+56$~Hz & \\
Binary on the equal-$\Ltil$ curve, $q\ge0.73$ & $\rho_{1\sigma}\gtrsim4\times10^4$ &\\
\botrule
\end{tabular}
\end{table*}

%=====================================================================
\section{Conclusions}\label{sec:conclusions}
%=====================================================================
We constructed a strange quark star and a neutron star with the same mass, quadrupolar tidal deformability and $f$-mode frequency in full general relativity. The quark matter lies inside the bulk stability window and supports $2\,\ms$, and the neutron star follows a calibrated analytic barotrope, with or without a realistic crust. The coincidence survives a 95~MeV strange-quark mass, an SLy envelope and a frozen-composition adiabatic index; its location is model dependent at the 10--20\% level in $\Lambda_2$, far beyond its numerical precision.

The coincidence reflects a general property rather than a fine-tuned accident. At fixed $(M,\Lambda_2)$ the $f$-mode frequencies of all our models that support $2\,\ms$ agree within about 0.5~Hz, and those of the matched pair within 0.25~Hz over $250\lesssim\Lambda_2\le1000$. This is below the 1.3~Hz linewidth of the mode and far below the projected measurement precision, so the $f$--$\Lambda_2$ relation cannot by itself separate strange stars from neutron stars. In binaries the two EoSs define a curve of equal $\Ltil$ on which the residual dephasing would require an SNR above $4\times10^4$ even with known masses. What does separate the two stars is their radius, which differs by 0.6--0.8~km once a crust is included, and their response at a second, larger mass.

All calculations assume cold, nonrotating, single-fluid stars. Elastic crusts, superfluidity, rotation and the conservative tidal response of each star would enter a quantitative interpretation of real data, and a full Bayesian analysis with marginalization over masses and spins would be needed for detector forecasts.

\backmatter
\begin{acknowledgments}
We are grateful to Feng-Li Lin, Jing-Yi Wu and Si-Man Wu for helpful discussions. Generative AI is applied in assisting the writing, literature research and coding. K.Z. (Hong Zhang) is supported by a classified fund of Shanghai city.
C.S.C. is partly supported by Taiwan's National Science and Technology Council through Grant No.~112-2112-M-032-012. C.V.F. acknowledges CNPq Project No.~304569/2022-4. C.H.L. acknowledges FAPESP Grant No.~2020/05238-9 and CNPq Grants No.~401565/2023-8, No.~409736/2025-2 and No.~305327/2023-2.
\end{acknowledgments}

\section*{Statements and Declarations}
\subsection*{Funding}
Funding sources and grant numbers are listed in the Acknowledgments.

\subsection*{Data availability}
Table~\ref{tab:czz} gives the nuclear equation-of-state coefficients. Tables~\ref{tab:coincidence}--\ref{tab:order} report the numerical stellar models, frequency comparisons and binary predictions. The observational inputs for GW170817 and GW190425 are taken from Refs.~\cite{binevent17,binevent19}.
% Authors: specify the access route to any additional underlying numerical files before submission.

\subsection*{Code availability}
The RelModPy code used for the relativistic perturbation equations is available at the revision identified in Ref.~\cite{relmod}. Appendix~\ref{app:numerics} describes the numerical driver and validation.
% Authors: add an access statement for the study-specific driver and analysis scripts.
% Authors: insert the agreed financial and non-financial competing interests statement.

% EPJ C continues figure and table numbering into the appendices.
\edef\lastmainfigure{\arabic{figure}}\edef\lastmaintable{\arabic{table}}
\begin{appendices}
\setcounter{figure}{\lastmainfigure}\setcounter{table}{\lastmaintable}
\renewcommand{\thefigure}{\arabic{figure}}\renewcommand{\thetable}{\arabic{table}}
\counterwithin*{equation}{section}

%=====================================================================
\section{Numerical methods and validation}\label{app:numerics}
%=====================================================================
\emph{Radiating modes.}---The relativistic calculation uses the polar interior right-hand side and the phase-amplitude Zerilli matching of the open-source RelModPy code~\cite{relmod}. Our driver integrates the coupled perturbations in $x=\ln(p_c/p)$ with two regular central columns and three surface columns satisfying $X=0$, matches all four interior variables, fixes the displacement normalization and uses the incoming exterior amplitude as the complex root function. The exterior contour and far boundary scale with the trial frequency. Every root is checked for positive damping, a nodeless radial displacement, the ratio of incoming to outgoing amplitude and its change under refinement. For the polytrope of Table~I of Lindblom, Mendell and Ipser~\cite{lmi} we obtain $\omega/\sqrt{M/R^3}=1.201618+3.9440\times10^{-4}i$ for $\ell=2$ and $1.586196+3.9667\times10^{-6}i$ for $\ell=3$, compared with $1.2016+3.95\times10^{-4}i$ and $1.5862+3.96\times10^{-6}i$. A strongly relativistic polytrope gives $M\omega=0.170844+6.1916\times10^{-5}i$ against $0.1708+6.193\times10^{-5}i$~\cite{kruger}. For SLy at $1.4\,\ms$ we find $f_2=1.9331$~kHz, within 0.06\% of an independent time-domain result~\cite{bernuzzi}.

\emph{Independent exterior and refinement.}---To check the small spectral differences we also integrate $Y=\dd\ln Z/\dd r_*$ from the vacuum exterior, $\dd Y/\dd r_*=V_Z-\omega^2-Y^2$, along the ray $r=R+ue^{-0.65i}$ on which the outgoing solution decays, starting from $Y=-i\omega$ at $u=45/{\rm Re}\,\omega$. The interior equations are shared with the production solver, but their matching null vector is obtained by singular-value decomposition. For both multipoles of the massless coincidence and of the two endpoint stars of Table~\ref{tab:triple}, this test reproduces all eight frequencies within $3.2\times10^{-8}$~Hz and the damping times within a fraction $2.4\times10^{-8}$. Tightening the interior and exterior integrations changes the coincident frequencies by less than $10^{-5}$~Hz and the inferred $\Lambda_*$ by at most $3\times10^{-6}$.

\emph{Cowling frequencies and static tides.}---The fixed-metric driver finds the lowest nodeless solutions by two-sided shooting. For the $K=100$, $\Gamma=2$ polytrope with $\rho_{0,c}=0.00128$ it gives $f_2^{\rm C}=1.88290$ and $f_3^{\rm C}=2.25646$~kHz, compared with the published 1.881 and 2.255~kHz~\cite{westernacher}. In the dilute bag limit it recovers $k_2=0.749997$ and $\omega_2^{\rm C}/\sqrt{M/R^3}=1.414213$; for the $n=1$ weak-field polytrope it gives $k_2=0.25990834$, compared with $(15-\pi^2)/(2\pi^2)=0.25990888$. A separately written code for the TOV equations, static tides and the Cowling problem, which shares no source with the production solver, reproduces the radii, $\Lambda_2$ and $M_{\max}$ of MAP, the bag, the $\Gamma=2$ polytrope and SLy in Table~\ref{tab:baseline}, the Cowling frequencies of the first three, the coincident mass and radius difference, and the values of $\Ltil$ in Sec.~\ref{sec:binary} to all quoted digits.

\emph{Moment of inertia and mismatch.}---To first order in the angular velocity $\Omega$ we solve $(r^4ju')'+4r^3j'u=0$ with $u=\Omega-\omega_{\rm drag}$ and $j=e^{-\Phi}\sqrt{1-2m/r}$, and set $I=R^4u'(R)/\{6[u(R)+Ru'(R)/3]\}$~\cite{hartle}. A volume integral for the angular momentum agrees to a fraction $10^{-7}$, the weak-field constant-density limit gives $I/(MR^2)=0.4000002$, and a separately written integrator, used for the crust-completed stars, reproduces the values of Sec.~\ref{sec:breaking} to $10^{-9}$. Overlaps use $2\times10^4$ logarithmically spaced frequencies with a numerical maximization over $t_c$; a second-order expansion in $\Delta\Psi$ agrees to better than 0.1\%.

%=====================================================================
\section{Comparison across equation-of-state shapes}\label{app:shapes}
%=====================================================================
In addition to CZZ, the bag model and SLy we consider three other shapes.
(i)~The Witten--Sakai--Sugimoto instanton-gas fit~\cite{liwss}, $\hat\epsilon=0.140\,\mathcal A^{0.571}\hat p^{0.429}+3.896\,\mathcal A^{-0.335}\hat p^{1.335}$ with $\mathcal A=1.8\times10^{-5}\ell^{-7}$, where $\ell$ is the asymptotic separation of the D8 and $\overline{\rm D8}$ branes in the units of Ref.~\cite{liwss}; we use $\ell^{-7}=10\,300$ and continue the fit to zero density.
(ii)~The strongly self-interacting quartic scalar fluid~\cite{colpi,flores}, $\hat\epsilon=3\hat p+\mathcal B\sqrt{\hat p}$ with $\mathcal B=0.08\lambda_4^{-1/2}(m/{\rm GeV})^2$ for a scalar of mass $m$ and quartic coupling $\lambda_4$; we use $\mathcal B=0.060$. The same functional form arises in a D3/D7 quark-matter model~\cite{hoyos}. We treat only the fluid limit, without field-gradient stresses or scalar radiation; scaling relations for the modes of such stars are discussed in Refs.~\cite{flores,shirke}.
(iii)~The relativistic polytrope $p=K\rho_0^2$, $\epsilon=\rho_0+p$, with $K=100$ in solar geometrized units~\cite{tooper}.
The WSS and quartic-fluid barotropes were the focus of earlier versions of this e-print. For SLy we invert the analytic fit of Ref.~\cite{hp} in $\log\rho$, use its analytic derivative, and below $10^5$~g\,cm$^{-3}$ continue the pressure as a power law with continuous value and slope. It gives $M_{\max}=2.048\,\ms$ at $R=9.981$~km, close to the tabulated 2.05 and 9.99~km~\cite{dh}. Table~\ref{tab:baseline} lists all shapes at their published normalizations.

\begin{table*}[tp]
\caption{Fiducial sequences at their published normalizations. $M_{\max}$ is the turning point of each sequence; the other entries refer to the $1.4\,\ms$ star. Damping times are full-GR amplitude e-folding times.}
\label{tab:baseline}
\centering\tablebodyfont
\begin{tabular}{@{}lrrrrrr@{}}
\toprule
EoS & $M_{\max}/\ms$ & $R$ (km) & $\Lambda_2$ & $f_2^{\rm GR}$ (kHz) & $f_2^{\rm C}$ (kHz) & $\tau_2$ (s)\\\midrule
WSS & 1.875 & 12.766 & 576.2 & 1.6750 & 2.0579 & 0.2556 \\
Quartic & 2.046 & 25.796 & 25\,984 & 0.6583 & 0.8220 & 2.0796 \\
Bag & 1.964 & 10.764 & 415.3 & 1.7998 & 2.3823 & 0.2206 \\
$\Gamma=2$ & 1.637 & 14.155 & 741.7 & 1.5780 & 1.8827 & 0.2935 \\
CZZ MAP & 2.444 & 11.329 & 516.9 & 1.7165 & 2.2494 & 0.2422 \\
CZZ median & 2.490 & 11.313 & 484.8 & 1.7407 & 2.2642 & 0.2355 \\
CZZ mean & 2.446 & 11.357 & 495.1 & 1.7328 & 2.2534 & 0.2377 \\
SLy & 2.048 & 11.701 & 297.0 & 1.9331 & 2.4086 & 0.1924\\
\botrule
\end{tabular}
\end{table*}

At fixed $M=1.4\,\ms$ and $\Lambda_2=512.1$ the seven rescaled shapes span 2.85~km in radius, 18.51~Hz in $f_2$ and 186.33~Hz in $f_3$ (Table~\ref{tab:order}). The three CZZ variants span 0.15~Hz in the radiating quadrupole and 15~Hz in Cowling. To identify the source of the octupolar spread we solve the static perturbation for $\ell=3$ with $y_\ell(0)=\ell$ and angular term $-\ell(\ell+1)/[r^2(1-2m/r)]$. The vacuum solutions are proportional to $P_3^2(r/M-1)$ and $Q_3^2(r/M-1)$, and matching gives $k_3$ and $\Lambda_3=2k_3/(15C^7)$~\cite{dn}. The vacuum-side jump of $y_\ell$ at a self-bound surface is the same as for $\ell=2$. At small compactness the solver gives $k_3=0.374998$ for the bag star and $0.106454$ for an $n=1$ polytrope, in agreement with the Newtonian values $3/8$ and $(105-10\pi^2)/(6\pi^2)$.

\begin{table*}[tp]
\caption{Matching the static tide at different multipole orders at $M=1.4\,\ms$. Each $\Delta f_\ell$ is the full-GR frequency minus the CZZ MAP frequency in the same group, in Hz.}
\label{tab:order}
\centering\tablebodyfont
\begin{tabular}{@{}lrrrrrr@{}}
\toprule
 & \multicolumn{3}{c}{Fixed $\Lambda_2=512.1$} & \multicolumn{3}{c}{Fixed $\Lambda_3=947.09$} \\
EoS & $\Lambda_3$ & $\Delta f_2$ & $\Delta f_3$ & $\Lambda_2$ & $\Delta f_2$ & $\Delta f_3$ \\\midrule
WSS & 1103.6 & $-0.99$ & $-66.53$ & 457.8 & 41.45 & $-12.04$ \\
Quartic & 1354.1 & $-18.37$ & $-182.66$ & 389.8 & 82.71 & $-63.23$ \\
Bag & 937.8 & $-0.10$ & 3.67 & 515.8 & $-2.81$ & 0.11 \\
$\Gamma=2$ & 1227.8 & $-7.47$ & $-124.95$ & 422.7 & 65.08 & $-34.31$ \\
CZZ MAP & 947.1 & 0 & 0 & 512.1 & 0 & 0 \\
CZZ median & 955.8 & 0.12 & $-3.47$ & 508.7 & 2.63 & $-0.15$ \\
CZZ mean & 957.2 & 0.14 & $-3.99$ & 508.2 & 3.04 & $-0.17$\\
\botrule
\end{tabular}
\end{table*}

Matching instead $\Lambda_3=947.09$ at the same mass reduces the $f_3$ range from 186.33 to 63.35~Hz and increases the $f_2$ range from 18.51 to 85.53~Hz. For the bag--MAP pair alone the octupolar separation falls from 3.67 to 0.11~Hz. Matching like multipoles thus absorbs most of the apparent spread, as expected from the diagonal $f$--Love relations~\cite{chan}; the residual, largest for the scalar fluid, reflects the shape of the barotrope.

SLy provides a nuclear shape beyond the two-power form. Matching it to $M=1.4\,\ms$ and $\Lambda_2=512.1$ requires $s=0.84524$ and gives $f_2=1.72042$ and $f_3=2.32705$~kHz, within 0.44~Hz of MAP in the quadrupole and 31.76~Hz in the octupole. Rematching at $\Lambda_3=947.09$ gives $f_3=2.35524$~kHz, 3.57~Hz below MAP. Both calculations use a fluid crust and the equilibrium adiabatic derivative.

The mass-support requirement matters for these ranges. Imposing $M_{\max}\ge2\,\ms$ on each rescaled sequence retains the bag model, the three CZZ shapes and SLy, among which the quadrupolar and octupolar frequency ranges are only 0.54 and 35.43~Hz. The rescaled quartic sequence, for example, has $M_{\max}\simeq1.41\,\ms$ at this target.

%=====================================================================
\section{Real-frequency gravitational-wave scattering}\label{app:scattering}
\begin{figure*}[tp]
\centering\includegraphics[width=\textwidth]{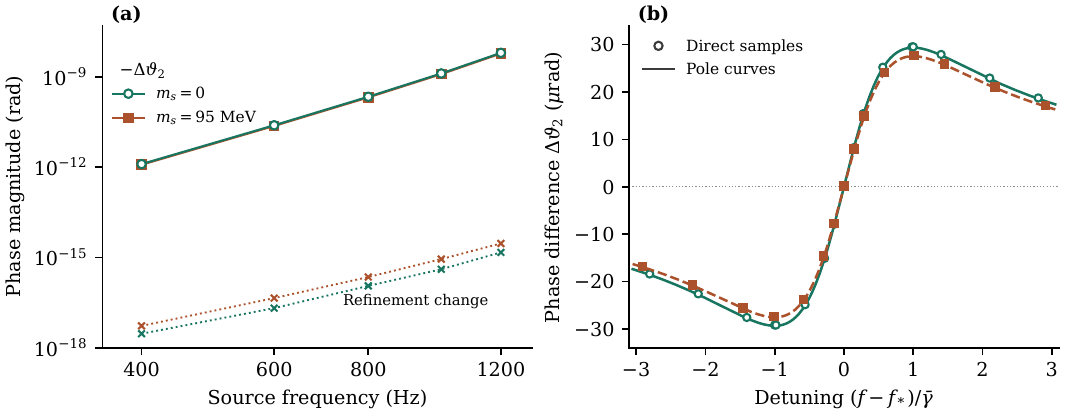}
\caption{Relative elastic scattering phase $\Delta\vartheta_2=\tfrac12\arg(S_{\rm bag}/S_{\rm MAP})$ for the two matched pairs. (a)~Below resonance $\Delta\vartheta_2<0$; symbols show $-\Delta\vartheta_2$ from the tighter integration and crosses the change from the baseline integration. (b)~Signed values against detuning from each pair's mean pole frequency $f_*$, in units of $\bar\gamma=\sqrt{\gamma_{\rm bag}\gamma_{\rm MAP}}$. Curves use the two complex poles of each pair, with no fitted background.}
\label{fig:scattering}
\end{figure*}

%=====================================================================
As a check of the pole structure we scatter a real-frequency even-parity $\ell=2$ gravitational wave off the massless coincident stars~\cite{stratton}. Each regular, free-surface interior solution is continued through vacuum to a common areal radius $r_m>\max(R_{\rm MAP},R_{\rm bag})$, where the real vector $(Z,\dd Z/\dd r_*)$ is decomposed on the outgoing Zerilli wave and its complex conjugate. With $S=A_{\rm out}/A_{\rm in}$ the phase normalization at $r_m$ is common to both stars and cancels from
\begin{equation}
 \Delta\vartheta(f)=\tfrac12\arg[S_{\rm bag}(f)/S_{\rm MAP}(f)].
\end{equation}
Flux conservation gives $|S|=1$, which holds to roundoff. Following the branch continuous from low frequency, $\Delta\vartheta$ grows from $-1.26\times10^{-12}$~rad at 400~Hz to $-6.25\times10^{-9}$~rad at 1200~Hz (Fig.~\ref{fig:scattering}). Varying the interior tolerances, matching radius and outgoing boundary together changes $\Delta\vartheta$ by less than $2\times10^{-9}$~rad, and an independent real-axis exterior confirms the phases within $6\times10^{-13}$~rad.

Near the common resonance $f_*$ each radiating pole of half-width $\gamma_i=(2\pi\tau_i)^{-1}$ contributes the factor $S_i^{\rm pole}=(f-f_{2,i}+i\gamma_i)/(f-f_{2,i}-i\gamma_i)$. At $f=f_*\mp0.713$~Hz the massless pair gives $\Delta\vartheta=(-2.919,+2.953)\times10^{-5}$~rad. Dividing the scattering ratio by the ratio of pole factors leaves $+1.65\times10^{-7}$~rad at both frequencies, so the independently computed damping times account for the phase offsets to 0.6\%. For the 95~MeV pair the residual is $+1.50\times10^{-7}$~rad. These phase differences are far from observable; an incident wave also poses a different boundary-value problem from tidal forcing by a companion~\cite{gaost,zhong}.

%=====================================================================

\end{appendices}


\begin{thebibliography}{99}
\bibitem{ak} N. Andersson and K. D. Kokkotas, Mon. Not. R. Astron. Soc. \textbf{299}, 1059 (1998), \href{https://arxiv.org/abs/gr-qc/9711088}{arXiv:gr-qc/9711088}.
\bibitem{chan} T. K. Chan, Y.-H. Sham, P. T. Leung, and L.-M. Lin, Phys. Rev. D \textbf{90}, 124023 (2014), \href{https://arxiv.org/abs/1408.3789}{arXiv:1408.3789}.
\bibitem{lau2010} H. K. Lau, P. T. Leung, and L. M. Lin, Astrophys. J. \textbf{714}, 1234 (2010), \href{https://arxiv.org/abs/0911.0131}{arXiv:0911.0131}.
\bibitem{chirenti2015} C. Chirenti, G. H. de Souza, and W. Kastaun, Phys. Rev. D \textbf{91}, 044034 (2015), \href{https://arxiv.org/abs/1501.02970}{arXiv:1501.02970}.
\bibitem{sotani2021} H. Sotani and B. Kumar, Phys. Rev. D \textbf{104}, 123002 (2021), \href{https://arxiv.org/abs/2109.08145}{arXiv:2109.08145}.
\bibitem{yagiyunes2017} K. Yagi and N. Yunes, Phys. Rep. \textbf{681}, 1 (2017), \href{https://arxiv.org/abs/1608.02582}{arXiv:1608.02582}.
\bibitem{zhao} T. Zhao and J. M. Lattimer, Phys. Rev. D \textbf{106}, 123002 (2022), \url{https://doi.org/10.1103/PhysRevD.106.123002}.
\bibitem{chu} I. G. Chu and C. A. Raithel, \href{https://arxiv.org/abs/2609.30143}{arXiv:2609.30143}.
\bibitem{witten} E. Witten, Phys. Rev. D \textbf{30}, 272 (1984).
\bibitem{afo} C. Alcock, E. Farhi, and A. Olinto, Astrophys. J. \textbf{310}, 261 (1986).
\bibitem{hzs} P. Haensel, J. L. Zdunik, and R. Schaeffer, Astron. Astrophys. \textbf{160}, 121 (1986).
\bibitem{postnikov} S. Postnikov, M. Prakash, and J. M. Lattimer, Phys. Rev. D \textbf{82}, 024016 (2010), \href{https://arxiv.org/abs/1004.5098}{arXiv:1004.5098}.
\bibitem{iloveself} T. K. Chan, A. P. O. Chan, and P. T. Leung, Phys. Rev. D \textbf{93}, 024033 (2016), \href{https://arxiv.org/abs/1511.08566}{arXiv:1511.08566}.
\bibitem{fukushima} K. Fukushima, J. Minamiguchi, and T. Uji, \href{https://arxiv.org/abs/2511.02236}{arXiv:2511.02236}.
\bibitem{pradhan2023} B. K. Pradhan, S. Shirke, and D. Chatterjee, \href{https://arxiv.org/abs/2311.15745}{arXiv:2311.15745}.
\bibitem{gao} D. Gao \textit{et al.}, Phys. Rev. D \textbf{114}, 043058 (2026), \url{https://doi.org/10.1103/w92f-58lx}, \href{https://arxiv.org/abs/2512.24703}{arXiv:2512.24703}.
\bibitem{raithel} C. A. Raithel and E. R. Most, Phys. Rev. D \textbf{108}, 023010 (2023), \url{https://doi.org/10.1103/PhysRevD.108.023010}.
\bibitem{twins} B. K. Pradhan, D. Chatterjee, and D. E. Alvarez-Castillo, Mon. Not. R. Astron. Soc. \textbf{531}, 4640 (2024), \url{https://doi.org/10.1093/mnras/stae1425}.
\bibitem{residual} G. Lioutas, A. Bauswein, and N. Stergioulas, Phys. Rev. D \textbf{104}, 043011 (2021), \url{https://doi.org/10.1103/PhysRevD.104.043011}.
\bibitem{lindblom92} L. Lindblom, Astrophys. J. \textbf{398}, 569 (1992).
\bibitem{czz} T.-S. Chen, X.-D. Zhou, and K. Zhang, Astrophys. J. \textbf{1009}, 227 (2026), \url{https://doi.org/10.3847/1538-4357/aea217}, \href{https://arxiv.org/abs/2607.00070}{arXiv:2607.00070}.
\bibitem{dh} F. Douchin and P. Haensel, Astron. Astrophys. \textbf{380}, 151 (2001), \href{https://arxiv.org/abs/astro-ph/0111092}{arXiv:astro-ph/0111092}.
\bibitem{hp} P. Haensel and A. Y. Potekhin, Astron. Astrophys. \textbf{428}, 191 (2004), \href{https://arxiv.org/abs/astro-ph/0408324}{arXiv:astro-ph/0408324}.
\bibitem{bag} P. Haensel, Acta Phys. Pol. B \textbf{18}, 739 (1987).
\bibitem{alford2005} M. Alford, M. Braby, M. Paris, and S. Reddy, Astrophys. J. \textbf{629}, 969 (2005).
\bibitem{auditweiss} S. Weissenborn, I. Sagert, G. Pagliara, M. Hempel, and J. Schaffner-Bielich, Astrophys. J. Lett. \textbf{740}, L14 (2011), \url{https://doi.org/10.1088/2041-8205/740/1/L14}.
\bibitem{farhi} E. Farhi and R. L. Jaffe, Phys. Rev. D \textbf{30}, 2379 (1984), \url{https://doi.org/10.1103/PhysRevD.30.2379}.
\bibitem{hinderer} T. Hinderer, Astrophys. J. \textbf{677}, 1216 (2008); \textbf{697}, 964(E) (2009), \href{https://arxiv.org/abs/0711.2420}{arXiv:0711.2420}.
\bibitem{dn} T. Damour and A. Nagar, Phys. Rev. D \textbf{80}, 084035 (2009), \url{https://doi.org/10.1103/PhysRevD.80.084035}.
\bibitem{ld} L. Lindblom and S. Detweiler, Astrophys. J. Suppl. Ser. \textbf{53}, 73 (1983).
\bibitem{dl} S. Detweiler and L. Lindblom, Astrophys. J. \textbf{292}, 12 (1985).
\bibitem{cowling} S. Yoshida and Y. Kojima, Mon. Not. R. Astron. Soc. \textbf{289}, 117 (1997), \href{https://arxiv.org/abs/gr-qc/9705081}{arXiv:gr-qc/9705081}.
\bibitem{fonseca} E. Fonseca \textit{et al.}, Astrophys. J. Lett. \textbf{915}, L12 (2021).
\bibitem{zhao2025} T. Zhao, P. B. Rau, A. Haber, S. P. Harris, C. Constantinou, and S. Han, \href{https://arxiv.org/abs/2504.12230}{arXiv:2504.12230}.
\bibitem{montefusco} G. Montefusco, M. Antonelli, and F. Gulminelli, \href{https://arxiv.org/abs/2410.08008}{arXiv:2410.08008}.
\bibitem{gaost} Y. Gao and J. Steinhoff, \href{https://arxiv.org/abs/2611.14243}{arXiv:2611.14243}.
\bibitem{binpratten} G. Pratten, P. Schmidt, and T. Hinderer, Nat. Commun. \textbf{11}, 2553 (2020), \url{https://doi.org/10.1038/s41467-020-15984-5}.
\bibitem{flanagan} E. E. Flanagan and T. Hinderer, Phys. Rev. D \textbf{77}, 021502(R) (2008).
\bibitem{binwade} L. Wade \textit{et al.}, Phys. Rev. D \textbf{89}, 103012 (2014), \url{https://doi.org/10.1103/PhysRevD.89.103012}.
\bibitem{raithelq} C. Raithel, F. \"Ozel, and D. Psaltis, Astrophys. J. Lett. \textbf{857}, L23 (2018), \href{https://arxiv.org/abs/1803.07687}{arXiv:1803.07687}.
\bibitem{yagibinary} K. Yagi and N. Yunes, Classical Quantum Gravity \textbf{33}, 13LT01 (2016), \url{https://doi.org/10.1088/0264-9381/33/13/13LT01}.
\bibitem{binevent17} B. P. Abbott \textit{et al.} (LIGO Scientific and Virgo Collaborations), Phys. Rev. X \textbf{9}, 011001 (2019), \url{https://doi.org/10.1103/PhysRevX.9.011001}.
\bibitem{binevent19} B. P. Abbott \textit{et al.} (LIGO Scientific and Virgo Collaborations), Astrophys. J. Lett. \textbf{892}, L3 (2020), \url{https://doi.org/10.3847/2041-8213/ab75f5}.
\bibitem{hinderer2016} T. Hinderer \textit{et al.}, Phys. Rev. Lett. \textbf{116}, 181101 (2016).
\bibitem{steinhoff2016} J. Steinhoff, T. Hinderer, A. Buonanno, and A. Taracchini, Phys. Rev. D \textbf{94}, 104028 (2016).
\bibitem{binfmtidal} P. Schmidt and T. Hinderer, Phys. Rev. D \textbf{100}, 021501(R) (2019), \href{https://arxiv.org/abs/1905.00818}{arXiv:1905.00818}.
\bibitem{lindblom2008} L. Lindblom, B. J. Owen, and D. A. Brown, Phys. Rev. D \textbf{78}, 124020 (2008).
\bibitem{keppel} D. Keppel and P. Ajith, Phys. Rev. D \textbf{82}, 122001 (2010), \href{https://arxiv.org/abs/1004.0284}{arXiv:1004.0284}.
\bibitem{bingw3g} N. Williams, G. Pratten, and P. Schmidt, Phys. Rev. D \textbf{105}, 123032 (2022), \url{https://doi.org/10.1103/PhysRevD.105.123032}.
\bibitem{gw170817} B. P. Abbott \textit{et al.} (LIGO Scientific and Virgo Collaborations), Phys. Rev. Lett. \textbf{119}, 161101 (2017).
\bibitem{apostol} T. Apostolidis \textit{et al.}, Phys. Rev. D \textbf{114}, 064087 (2026), \url{https://doi.org/10.1103/37rp-dwhy}.
\bibitem{hegade2026} A. Hegade K. R., K. J. Kwon, T. Venumadhav, H. Yu, and N. Yunes, \href{https://arxiv.org/abs/2605.08569}{arXiv:2605.08569}.
\bibitem{ilove} K. Yagi and N. Yunes, Science \textbf{341}, 365 (2013), \href{https://arxiv.org/abs/1302.4499}{arXiv:1302.4499}.
\bibitem{kramer} M. Kramer \textit{et al.}, Phys. Rev. X \textbf{11}, 041050 (2021), \url{https://doi.org/10.1103/PhysRevX.11.041050}.
\bibitem{relmod} F. Gittins, RelModPy v0.2.0, \href{https://github.com/fgittins/RelModPy/tree/fe3abd1c787ce741f035c4b63fbec2c305b56932}{commit fe3abd1} (MIT license).
\bibitem{lmi} L. Lindblom, G. Mendell, and J. R. Ipser, Phys. Rev. D \textbf{56}, 2118 (1997), \url{https://doi.org/10.1103/PhysRevD.56.2118}.
\bibitem{kruger} C. J. Kr\"uger, Ph.D. thesis, University of Southampton, 2015, \href{https://eprints.soton.ac.uk/384187/1/ckrueger_thesis.pdf}{Table 3.3}.
\bibitem{bernuzzi} S. Bernuzzi and A. Nagar, Phys. Rev. D \textbf{78}, 024024 (2008), \href{https://arxiv.org/abs/0803.3804}{arXiv:0803.3804}.
\bibitem{westernacher} J. R. Westernacher-Schneider \textit{et al.}, Phys. Rev. D \textbf{100}, 123009 (2019), \url{https://doi.org/10.1103/PhysRevD.100.123009}.
\bibitem{hartle} J. B. Hartle, Astrophys. J. \textbf{150}, 1005 (1967), \url{https://doi.org/10.1086/149400}.
\bibitem{liwss} W. Li, J.-Y. Wu, and K. Zhang, Results Phys. \textbf{64}, 107893 (2024), \href{https://arxiv.org/abs/2403.20240}{arXiv:2403.20240}.
\bibitem{colpi} M. Colpi, S. L. Shapiro, and I. Wasserman, Phys. Rev. Lett. \textbf{57}, 2485 (1986), \url{https://doi.org/10.1103/PhysRevLett.57.2485}.
\bibitem{flores} C. V\'asquez Flores, A. Parisi, C.-S. Chen, and G. Lugones, J. Cosmol. Astropart. Phys. 06 (2019) 051, \href{https://arxiv.org/abs/1901.07157}{arXiv:1901.07157}.
\bibitem{hoyos} C. Hoyos, D. Rodr\'iguez Fern\'andez, N. Jokela, and A. Vuorinen, Phys. Rev. Lett. \textbf{117}, 032501 (2016), \href{https://arxiv.org/abs/1603.02943}{arXiv:1603.02943}.
\bibitem{shirke} S. Shirke, B. K. Pradhan, D. Chatterjee, L. Sagunski, and J. Schaffner-Bielich, \href{https://arxiv.org/abs/2502.04059}{arXiv:2502.04059}.
\bibitem{tooper} R. F. Tooper, Astrophys. J. \textbf{142}, 1541 (1965).
\bibitem{stratton} T. Stratton and S. R. Dolan, Phys. Rev. D \textbf{100}, 024007 (2019), \url{https://doi.org/10.1103/PhysRevD.100.024007}.
\bibitem{zhong} B. Zhong, Z. Miao, Z. Pan, and H. Yang, \href{https://arxiv.org/abs/2611.22417}{arXiv:2611.22417}.
\end{thebibliography}
\end{document}